\documentclass[12pt,preprint]{aastex}

\usepackage{amsmath,amssymb}

\newcommand{\bfm}[1]{{\mbox{\boldmath $#1$}}}
\begin{document}

\title{The relation between the two-point and the three-point
correlation functions in the non-linear gravitational clustering regime}

\author{Hiroko Koyama$^{1,2}$ and Taihei Yano$^1$}
\affil{$^{1}$National Astronomical Observatory, Mitaka, Tokyo 181--8588, Japan}
\affil{$^{2}$Department of Physics, Waseda University, Shinjuku, Tokyo 169--8555, Japan}

\email{koyama@gravity.phys.waseda.ac.jp, yano.t@nao.ac.jp}

\begin{abstract}
The connection between the two-point and the three-point correlation 
functions in the non-linear gravitational clustering regime is studied.
Under a scaling hypothesis, we find that the three-point correlation
function, $\zeta$, obeys the scaling law 
$\zeta\propto \xi^{\frac{3m+4w-2\epsilon}{2m+2w}}$ in the nonlinear regime, where
$\xi$, $m$, $w$, and $\epsilon$ are the two-point correlation function,
the power index of the power spectrum in the nonlinear regime, the
number of spatial dimensions, and the power index of the phase
correlations, respectively.
The new formula reveals the origin of the power index of the 
three-point correlation function. 
We also obtain the theoretical condition for which the 
``hierarchical form'' $\zeta\propto\xi^2$ is reproduced.

\end{abstract}

\section{Introduction}
The formation of large-scale structures in the Universe is
one of the most important and interesting problems in cosmology.
It is generally believed that these structures have developed by a process of gravitational instability from small initial fluctuations in the density of a largely homogeneous early Universe. 
Hence it is very important to clarify the physical mechanism of 
the evolution of density fluctuations. 
Here we consider the density fluctuations of collisionless particles, such as dark matter, because our interest is concentrated on the effects of self-gravity.

The inflationary scenario predicts  
a random Gaussian field possessing the properties of statistical
homogeneity and isotropy as the primordial fluctuations
\citep{gut81,gut82,bar86}.
The statistical properties of the Gaussian field are completely
characterized by the two-point correlation function
(hereafter 2PCF) or the power spectrum, since higher order correlations vanish, while only the 2PCF has a certain value \citep{adl81}.
However, non-linear gravitational clustering gives rise to 
non-zero values of the higher-order correlations, 
even if the primordial density fluctuations were set Gaussian. 
This is reasonable, since phases between Fourier modes in the bispectrum
are considered to be distributed non-randomly but correlated in some ways
in the non-linear regime, where galaxies or galaxy groups are 
strongly clustered around each other \citep{sch91,col00,chi01,chi02,coles,wat03b,matsubara03}.
Hence the higher order statistics, like the three-point correlation function (hereafter 3PCF), become essential
when the evolution of the density fields is governed by non-linear gravitational clustering.
So it is quite important for understanding 
non-linear evolution to investigate the properties of the 3PCF.

Generally, it is hard to deal with the exact formula of the 3PCF
both theoretically and numerically, since it
is much more lengthy and complicated than 2PCF. Historically, however, 
\citet{pg75,gp77} had proposed a simple assumption: the 3PCF
can be expressed by the product of the 2PCF as
\begin{equation}
\zeta_{abc}=Q[\xi_{ab}\xi_{bc}+\xi_{bc}\xi_{ca}+\xi_{ca}\xi_{ab}],
\label{assumption}
\end{equation}
where $\zeta$, $\xi$, and $Q$ are the 3PCF, the 2PCF, and a certain constant, respectively.
The expression (\ref{assumption}) or $\zeta \propto \xi^2$ is often
called the ``hierarchical form'' \citep{f84}.
The data from the Zwicky and Shane-Wirtanen catalogs are in good
agreement
with this assumption when we choose $Q\simeq 0.85$ for Zwicky and $Q\simeq
1.24$ for Shane-Wirtanen \citep{pg75,gp77}.

However, there is a serious fault with the proposal (\ref{assumption}).
It has no theoretical ground, 
although some observational or numerical data may be well-fitted.
One explanation
 may be that Eq. (\ref{assumption}) for $\zeta$ with $Q=1$ coincides with the Kirkwood superposition approximation
which is familiar in liquid physics and turbulence theory, 
if $\xi_{ab}\xi_{bc}\xi_{ca}\ll 1$ \citep{dp77,ichimaru73}.
However, this approximation is not appropriate
for the non-linear regime, since the distribution of the gravitational sources
is strongly correlated, $\xi\gg 1$, in such regimes.
Hence this resemblance is helpless to compensate for the theoretical defect.

Indeed, there were many theoretical investigations of the 3PCF 
\citep{f84,rf92}.
Nevertheless most of
the analyses have been based on the ``hierarchical form'' assumption.
The coefficient $Q$ under the ``hierarchical form'' 
was estimated by assuming self-similarity \citep{f84}.
The BBGKY hierarchy was analyzed 
by assuming the ``hierarchical form'' as well as self-similarity \citep{rf92}.
In the first place, however, there is no reason why the 3PCF
should necessarily depend only on the second power of the 2PCF,
even if we accept the self-similarity. 
Indeed, \citet{yg97} suggests the possible existence of solutions that do not satisfy the ``hierarchical form'' on the basis of the BBGKY equations.
Therefore an investigation of the physical adequacy of the  ``hierarchical form'' itself
and more reliable and useful formulae
for the 3PCF are pressingly required. 

In this paper we study the connection between the 2PCF and the 3PCF
in the non-linear gravitational clustering regime. 
We analyze these functions using a scaling hypothesis.
It is expected that the time evolution of the statistics obeys also some 
self-similar rules fundamentally, since gravity is scale free \citep{dp77,f84,rf92,yg98}. 
Many authors have investigated the self-similarity of the 2PCF or the power spectrum.
Recent works using N-body simulations \citep{cbh1996,cp1997,jmw1995}
have shown that self-similar evolution of the power spectrum can be satisfied when the initial power spectrum is 
scale free, and the power spectrum in the non-linear regime obeys some scaling laws.
The scaling hypothesis itself is very familiar. 
However, the unique proposition in this paper is that 
we apply the scaling hypothesis to the bispectrum as well as to the power spectrum, and we never assume the ``hierarchical form''.
Consequently, we succeed in deriving a new formula for the 3PCF, focusing on the power indices.
The relevance to the ``hierarchical form'' is also discussed.

Our paper is organized as follows.
In Sec.\ref{sec:general} we introduce the general formula of the 2PCF and the 3PCF.
We obtain in Sec.\ref{sec:correlation} a new relationship
between these functions the under the scaling hypothesis.
In Sec.\ref{sec:simulation} we calculate the 2PCF and the 3PCF 
in the non-linear regime
in the one-dimensional Einstein-de Sitter model numerically,
in order to confirm the analytical results.
The final section is devoted to summary.

\section{General formula for the correlation functions}
\label{sec:general}

In this section we consider the 2PCF and the 3PCF.
Let us denote by $\bar{\rho}$ the mean density 
and take $\rho(\bfm{x})$
to be the density at a point specified by the position vector $\bfm{x}$ with
respect to some arbitrary origin. As usual, we define the fluctuation to
be
\begin{equation}
 \delta(\bfm{x})=\frac{\rho(\bfm{x})-\bar{\rho}}{\bar{\rho}}.
\end{equation}
We assume this to be expressible as a Fourier series:
\begin{equation}
\delta _{\bfm{k}}=\frac{1}{L^w}\int _{} \delta (\bfm{x}) e^{i\bfm{k}\cdot\bfm{x}}d^wx,
\end{equation}
and
\begin{equation}
\delta(\bfm{x})=\frac{L^w}{(2\pi)^w}
\int \delta _{\bfm{k}}e^{-i\bfm{k}\cdot\bfm{x}}d^wk,
\end{equation}
where $\delta_{\bfm{k}}$, $w$ and $L$ are a Fourier amplitude,
the number of the spatial dimensions, and the length of the side of the box, respectively.
The 2PCF is defined in terms of the density
fluctuation by
\begin{eqnarray}
 \xi(\bfm{r})
&=&\left\langle\left\langle
\delta(\bfm{x})\delta(\bfm{x} +\bfm{r})\right\rangle_{\bfm{x}}
\right\rangle_{\rm }
\nonumber\\
&=&
\left\langle\left\langle
\frac{L^{2w}}{(2\pi)^{2w}}
\iint \delta_{\bfm{k}} \delta_{\bfm{k'}}
e^{-i(\bfm{k}+\bfm{k'})}\cdot \bfm{x}e^{-i\bfm{k}\cdot\bfm{r}}
d^{w}kd^{w}k'\right\rangle_{\bfm{x}}\right\rangle_{\rm},
\label{eq:xi-bracket}
\end{eqnarray}
where the angular brackets 
in Eq.(\ref{eq:xi-bracket}) are expectation values, formally
denoting an average over the probability distribution of $\delta(\bfm{x})$.
The first average, $\langle\quad\rangle_{\bfm{x}}$, means simply a volume
normalization over a sufficiently large patch of the Universe. The second
average, $\langle\quad\rangle$, is interpreted as that across independent samples in many Universes.
Applying this rule and performing the volume
integration gives
\begin{eqnarray}
 \xi(\bfm{r})
&=&\frac{L^{w}}{(2\pi)^{w}}
\iint
\langle\delta_{\bfm{k}}\delta_{\bfm{k'}}\rangle_{\rm }
\delta_D(\bfm{k}+\bfm{k'})e^{-i\bfm{k}\cdot\bfm{r}}
d^{w}k'd^{w}k\nonumber\\
&\equiv& \frac{L^{w}}{(2\pi)^{w}}
\int P(\bfm{k})e^{-i\bfm{k}\cdot\bfm{r}}
d^{w}k,
\label{eq:xiP}
\end{eqnarray}
where $\delta_D$ is the Dirac delta-function and 
$P(\bfm{k})$ is the power spectrum. 
Similarly, it is possible to calculate the 3PCF, which is defined as
\begin{eqnarray}
 \zeta(\bfm{r},\bfm{s})
&=&\langle\langle\delta(\bfm{x})\delta(\bfm{x}+\bfm{r})\delta(\bfm{x}+\bfm{s})
\rangle_{\bfm{x}}\rangle
\nonumber\\
&=&
\left\langle\left\langle\frac{L^{3w}}{(2\pi)^{3w}}
\iiint \delta_{\bfm{k}} \delta_{\bfm{k'}}\delta_{\bfm{k''}}
e^{-i(\bfm{k}+\bfm{k'}+\bfm{k''})\cdot \bfm{x}}e^{-i(\bfm{k}\cdot\bfm{r}+\bfm{k'}
\cdot\bfm{s})}
d^{w}kd^{w}k'd^{w}k''\right\rangle_{\bfm{x}}\right\rangle.
\end{eqnarray}
The volume integration gives 
\begin{eqnarray}
 \zeta(\bfm{r},\bfm{s})
&=&
\frac{L^{2w}}{(2\pi)^{2w}}\iiint 
\langle\delta_{\bfm{k}} \delta_{\bfm{k'}}\delta_{\bfm{k''}}\rangle_{\rm}
\delta_D(\bfm{k}+\bfm{k'}+\bfm{k''})
e^{-i(\bfm{k}\cdot\bfm{r}+\bfm{k'}\cdot\bfm{s})}
d^{w}kd^{w}k'd^{w}k''
\nonumber\\
&=&
\frac{L^{2w}}{(2\pi)^{2w}}\iint 
\langle\delta_{\bfm{k}} \delta_{\bfm{k'}}\delta_{-\bfm{k}-\bfm{k'}}\rangle_{\rm }
e^{-i(\bfm{k}\cdot\bfm{r}+\bfm{k'}\cdot\bfm{s})}
d^{w}kd^{w}k'
\nonumber\\
&\equiv&
\frac{L^{2w}}{(2\pi)^{2w}}\iint 
B(\bfm{k},\bfm{k'})e^{-i(\bfm{k}\cdot\bfm{r}+\bfm{k'}\cdot\bfm{s})}
d^{w}kd^{w}k',
\label{eq:zetaB}
\end{eqnarray}
where $B(\bfm{k},\bfm{k'})$ is the bispectrum \citep{f84}. 
The volume integration gives rise to
the triangular constraint by the Dirac delta-function, 
which is similar to
Eq.(\ref{eq:xiP}) (see also \citet{f84,coles,matsubara03}).

\section{The connection between the 2PCF and the 3PCF 
under the scaling hypothesis}
\label{sec:correlation}
In this section we derive the connection between the 2PCF and the 3PCF,
under the scaling hypothesis. This assumption is adequate, since we
consider that fluctuations evolve purely due to gravity.
First, we assume that the power spectrum in the non-linear regime
obeys a power law as 
\begin{eqnarray}
 P(k)&=&\langle |\delta_k|^2\rangle \nonumber\\
&\propto& k^m,
\label{eq:power}
\end{eqnarray}
where the index is a value in the non-linear regime.
Substituting Eq.(\ref{eq:power}) into Eq.(\ref{eq:xiP}) gives
\begin{eqnarray}
\xi(r)&\propto&
\int 
k^{m}
d^{w}k
\nonumber\\
&\propto& r^{-(m+w)}.
\label{eq:xi-kekka}
\end{eqnarray}
Next, the bispectrum in Eq.(\ref{eq:zetaB}) is
decomposed by
\begin{equation}
B(k,k')= \left\langle |\delta_k||\delta_{k'}||\delta_{-k-k'}|
e^{i(\phi_k+\phi_{k'}+\phi_{-k-k'})}\right\rangle_{\rm },
\label{bispectrum}
\end{equation}
where $\phi_{k}$ is the Fourier phase of the modes with wave number
$k$. Now we consider the Fourier phases in Eq.(\ref{bispectrum})
in the non-linear regime.
Let us consider that the initial fluctuations are a random Gaussian field whose phases are distributed randomly.
Induced by the  gravitational instability, however, 
the density fluctuations become stronger and stronger,
which means 
the phases between Fourier modes become distributed
non-randomly \citep{sch91,col00,chi01,chi02,coles,wat03b}.
That is, the Fourier phases are also correlated 
due to the gravitational clustering.
It is not difficult to imagine that the phases synchronize
in the case
of the gravitational collapse, like a pancake, for example.

Generally, the 3PCF or the bispectrum depends on 
the shape of the configuration.
If we fix the shape of the triangle, 
however, the 3PCF and the bispectrum are determined by one parameter.
Here we assume the bispectrum in the strong non-linear regime
obeys a scaling law as well as the power spectrum.
Hence 
the expectation value of the Fourier phase in the bispectrum 
(\ref{bispectrum}) is described as
\begin{equation}
 \langle e^{i(\phi_{kt}+\phi_{k't}+\phi_{-kt-k't})}\rangle
\propto t^{-\alpha} 
\langle e^{i(\phi_{k}+\phi_{k'}+\phi_{-k-k'})}\rangle
\label{Fourier-phase}
\end{equation}
in the strongly non-linear regime.
The scaling hypothesis for the Fourier phases 
corresponds to the following two characteristic possibilities physically:
first, the Fourier phases remain correlated, and 
the expectation value converges to some non-zero constant
value through non-linear gravitational clustering.
In this case the value of $\alpha$ in Eq.(\ref{Fourier-phase}) is zero.
Second, the Fourier phase correlations are
finally relaxed due to non-linear gravitational clustering,
although the phases may be correlated once by gravitational instability.
In this case the expectation value of the Fourier-phase correlations
converges to zero, and the value of $\alpha$ in
Eq.(\ref{Fourier-phase}) is some finite
value that depends on the degree of convergence.
That is, the constant $\alpha$ should be a non-negative value.
Similarly, other terms in Eq.(\ref{bispectrum}) 
are also assumed to obey some scaling laws (For a detailed description, see Appendix \ref{sec:ap}). 

From Appendix \ref{sec:ap}, the 3PCF is reduced to 
\begin{eqnarray}
 \zeta(r,s)
&\propto&
\iint B(k,k')d^{w}kd^{w}k'
\nonumber\\
&\propto&
t^{-(\frac{3}{2}m+2w-\alpha)}
\iint
\langle |\delta_{tk}|\rangle \langle |\delta_{tk'}|\rangle
\langle |\delta_{-tk-tk'}|\rangle
\langle e^{i(\phi_{tk}+\phi_{tk'}+\phi_{-tk-tk'})} \rangle
d^{w}(tk)d^{w}(tk')
\nonumber\\
&&+t^{-(\frac{3}{2}m+2w-\beta-\gamma_{4y})}
\iint
\langle |\delta_{tk}|\rangle \langle |\delta_{tk'}|\rangle
\langle |\delta_{-tk-tk'}|\rangle
\langle \delta e^{i(\phi_{tk}+\phi_{tk'}+\phi_{-tk-tk'})} \rangle
d^{w}(tk)d^{w}(tk')
\nonumber\\
&&+\cdots.
\end{eqnarray}
Where $\beta$ and $\gamma$ are the scaling factors of variance and other
constants. 
It is noting that each value of these factors should be non-negative
so that the 3PCF should not diverge.
Then the 3PCF can be approximated by
\begin{eqnarray}
\zeta(r,s)
&\propto&
t^{-(\frac{3}{2}m+2w-\epsilon)}
 \zeta\left(\frac{r}{t},\frac{s}{t}\right),
\label{eq:zeta-kekka0}
\end{eqnarray}
that is,
\begin{equation}
 \zeta(r)\propto r^{-(\frac{3}{2}m+2w-\epsilon)},
\label{eq:zeta-kekka}
\end{equation}
where $\epsilon$ is defined as
\begin{equation}
 \epsilon \equiv 
min(\alpha,\> \beta+\gamma_{2y},\> \beta+\gamma_{3y},\>\beta+\gamma_{4y}).
\label{eq:delta}
\end{equation}
From Eqs.(\ref{eq:xi-kekka}) and (\ref{eq:zeta-kekka}),
we obtain a new expression to connect between $\zeta$ and $\xi$ as
\begin{equation}
 \zeta\propto \xi^{\frac{3m+4w-2\epsilon}{2m+2w}}.
\label{eq:results}
\end{equation}
The proportional expression (\ref{eq:results}) states that 
the 3PCF is proportional to the power of the 2PCF.
The power index depends on the power index of the
power spectrum in the non-linear regime, $m$, and the scaling factors of the bispectrum, $\epsilon$.
The formula clarifies 
the source of the power index of the 3PCF under the scaling hypothesis.

Now we consider the case $\alpha\ne 0$,
where the phase the phase correlations get loose
by the  non-linear gravitational clustering,
although they are tied tight once. 
Still the information of the convergence survives in 
the power index of the formula (\ref{eq:results}) .
We can know the power index of the 3PCF by
calculating the power index of the power spectrum and 
the phase correlation in the bispectrum (11).
In particular, if the equality $\epsilon=-m/2$ is satisfied, 
the ``hierarchical form'', $\zeta\propto\xi^2$, can be reproduced.
For example, we consider the case  $\xi\propto r^{-1.8}$
[Davies\&Peebles 1977].
In this case, the index $m$ is equal to $-1.2$, and
our formula agrees to the ``hierarchical form'',
if $\epsilon=0.6$.

Next, we consider the case where the Fourier phases keep correlated through
the non-linear effects. In this case 
the expectation value of the phase correlations converges to a finite constant
value,  $\alpha =0$. 
Then the power index of the formula (\ref{eq:results})  depends only on  
the index of the power spectrum and the number of the spatial
dimension,
\begin{eqnarray}
\zeta\propto \xi^{\frac{3m+4w}{2m+2w}}.
\label{formula-alphazero}
\end{eqnarray}
Accordingly, the 3PCF can be described as the closed form only by the
2PCF. As a typical example, it is expected that the relation is realized at the moment of
the gravitational collapse into a pancake, since the phases are
synchronized strongly.
In addition, if the power index of the power spectrum is zero, $m=0$,
the 3PCF is proportional to the second power of the
2PCF. Recall that the index of the power spectrum is
around zero in the transitional region from
the linear regime to the non-linear regime, if the power index of the
power spectrum of the initial random Gaussian field is not negative value.
Considering these facts, we suppose 
the ``hierarchical form'' might be also realized accidentally
around such quasi-linear scales.

\section{Numerical test in the one-dimensional model}
\label{sec:simulation}
In this section we demonstrate the numerical simulation by the
one-dimensional model to confirm our analytical expression.
This model enables us to simulate with high accuracy, since 
the numerical errors of the algorithm include only round-off errors.
Here we investigate the time evolution of a plane--symmetric density 
perturbation in an Einstein--de Sitter universe model by using a
semi--analytical method.
In the one--dimensional system many plane--parallel sheets move only 
in perpendicular direction to the surface of these sheets.
When two sheets cross, they are allowed to pass through each other freely.
In this sheet system, there is an exact solution
until two sheets cross over as follows
\citep{z1970,sz1972,drs1973,zc1980,yg98}:
\begin{eqnarray}  
x &=& q + B_1(t)S_1(q) + B_2(t)S_2(q), \nonumber \\
v &=& \dot{B_1}(t)S_1(q) + \dot{B_2}(t)S_2(q)\;\;, 
\label{1}
\end{eqnarray}  
where $q$ and $x$ are the Lagrangian and the (comoving) Eulerian coordinates, 
respectively.
Here, $S_1(q)$, and $S_2(q)$ are arbitrary functions of $q$. 
$B_1(t)$, and $B_2(t)$ are the growing mode and the decaying mode of linear 
perturbation solutions, respectively.
We are considering the Einstein--de Sitter background universe model, i.e., 
$B_1(t)=a$ and $B_2(t)=a^{-\frac{3}{2}}$,
where $a$ is the scale factor. The velocity $v$ is the peculiar--velocity
normalized by the scale factor.
We can compute the crossing time of all neighboring pairs of sheets.
We use the shortest of these crossing times as a time step.
Then, we can compute the new positions and velocities for all sheets 
at this crossing time.
After two sheets cross, we exchange the velocities of the two sheets that
just crossed.
Then we obtain again $S_1(q), S_2(q)$, and therefore the exact solutions
as follows:
\begin{eqnarray}  
S_1(q)=\frac{3}{5}a^{-1}(x-q)+\frac{2}{5}\dot{a}^{-1}v, \nonumber \\
S_2(q)=\frac{2}{5}a^{\frac{3}{2}}(x-q)
 -\frac{2}{5}\dot{a}^{-1}a^{\frac{5}{2}}v\;\;.
\end{eqnarray}  
These new exact solutions can be used until again two sheets cross.
In this way we obtain the exact loci of the sheets 
by coupling these solutions.
This semi--analytic method has good accuracy, because 
we connect exact solutions.

\begin{figure}
\plotone{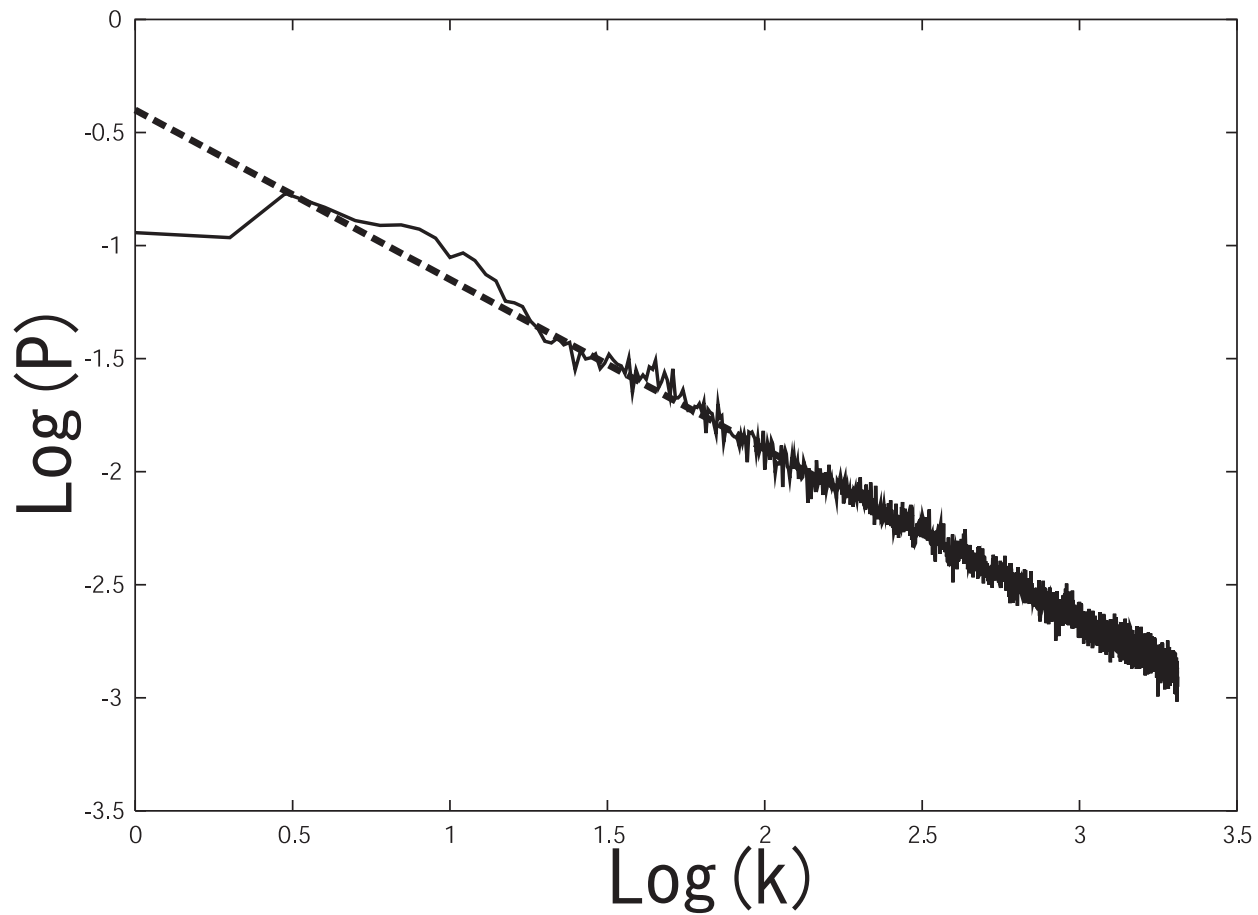}
 \caption{Power spectrum $P(k)$. We have chosen the initial power spectrum as
 the power law with index $n=1$. The power index of the power
 spectrum in the non-linear regime is fitted to $-0.75$.\label{fig1}}  
\end{figure}

\begin{figure}
\plotone{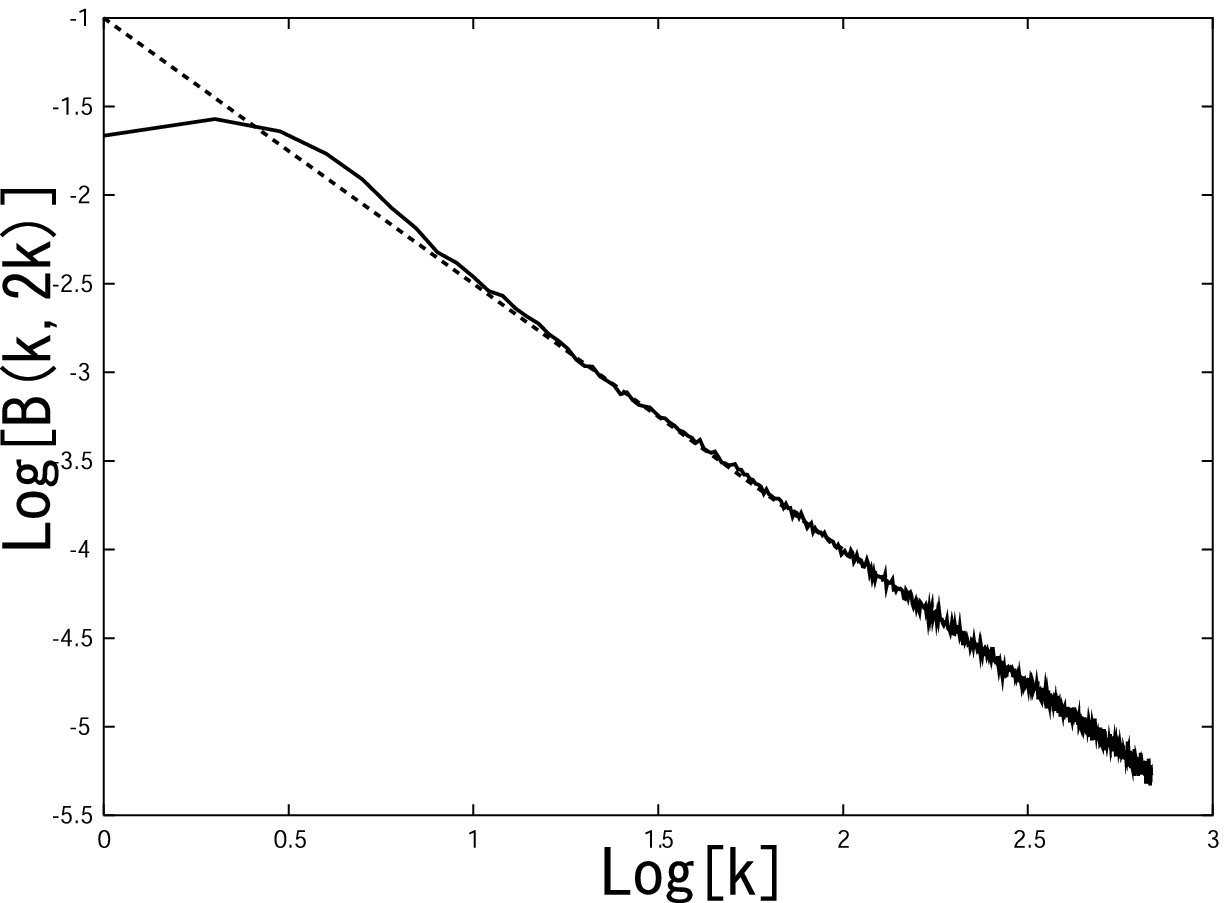}
 \caption{The bispectrum in the nonlinear regime when the power spectrum is
expressed as in Fig.1. The dashed line shows a power law with a power
index of $-1.5$.\label{fig2}}
\end{figure}

\begin{figure}
\plotone{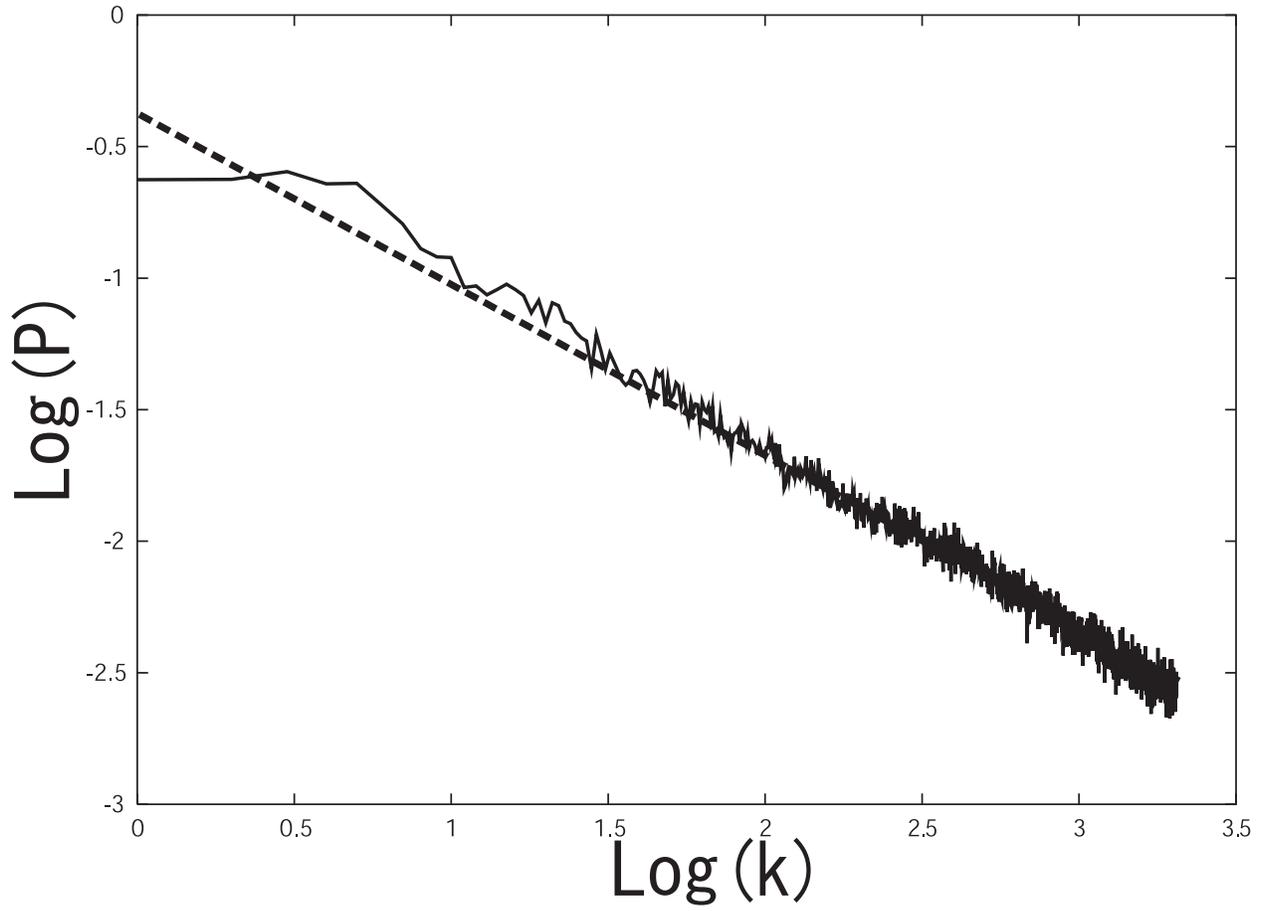}
 \caption{The same as Figure \ref{fig1}, but here the initial power index is $2$.
The power index of the power spectrum in the non-linear regime is fitted 
by $-0.65$.\label{fig3}}  
\end{figure}

\begin{figure}
\plotone{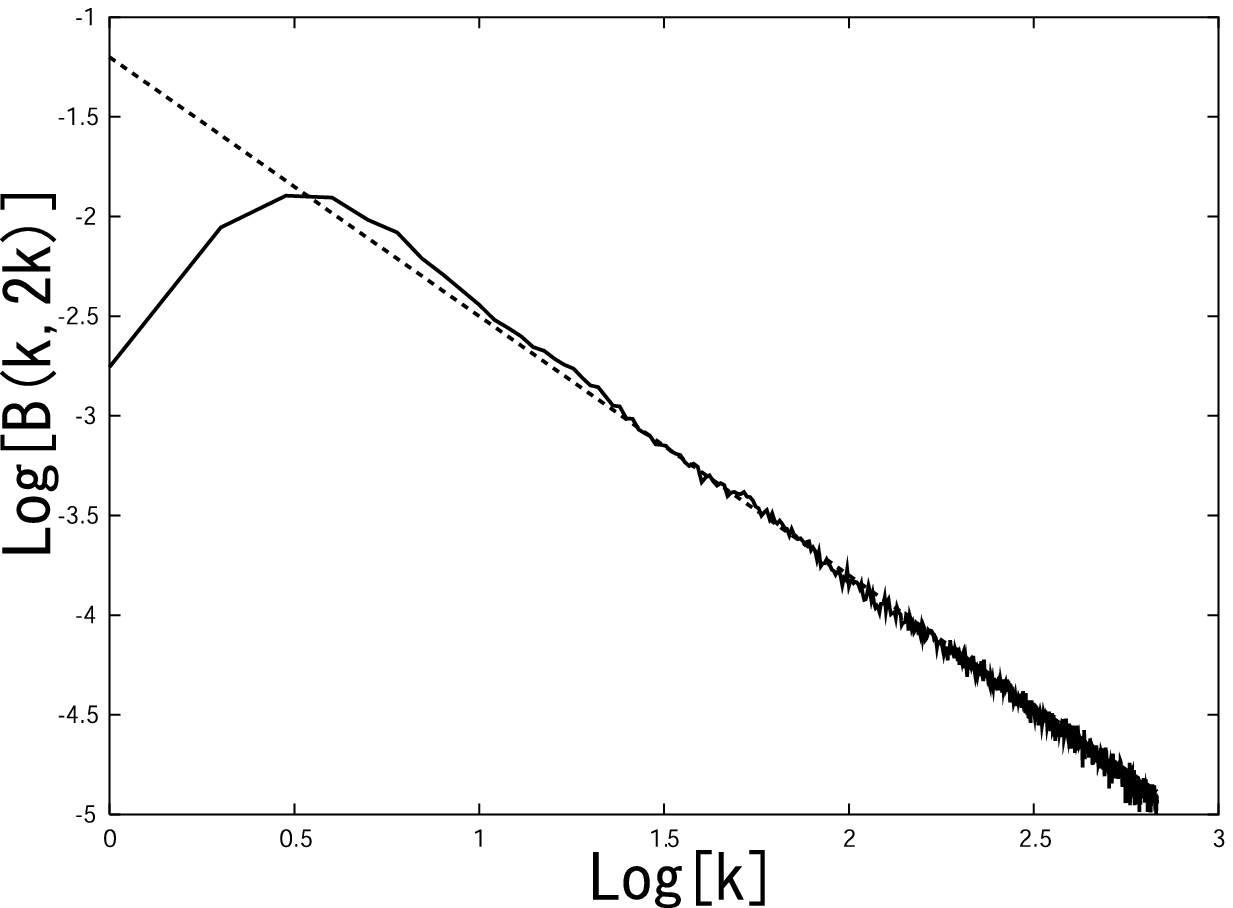}
 \caption{
The bispectrum in the nonlinear regime when the power spectrum is
expressed as in Figure \ref{fig3}. The dashed line shows a power law with a power
index of $-1.3$.\label{fig4}}  
\end{figure}

In our calculation we are going to use $2^{12}$ sheets.
Periodic boundary conditions are fixed at a length of $2$. 
Here we consider the time evolution from a random-Gaussian initial condition with the index of the power spectrum given by $1$ or $2$. 
The shape of the triangle is chosen as $r=2s$.
In addition, we take the ensemble average over 100 samples.
Figure \ref{fig1} shows the power spectrum in the nonlinear
gravitational regime where we have chosen the initial power index
$1$.  It can be fitted by $P(k)\propto k^{-0.75}$,
which agrees with \citet{yg98}. From the Fourier transform,
the 2PCF obeys $\xi(r)\propto r^{-0.25}$.
Figure \ref{fig2} shows the bispectrum in the nonlinear regime when the power spectrum is expressed as in Figure \ref{fig1}. 
It is obvious that the bispectrum
is well fitted by $B(k,2k)\propto k^{-1.5}$. 
Similarly, Figure \ref{fig4} shows the bispectrum in the nonlinear regime when the power
spectrum is expressed as in Figure \ref{fig3}. It is obvious that the bispectrum
is well fitted by $B(k,2k)\propto k^{-1.3}$. 
These value agree with $B(k,2k)\propto k^{2m}$,
which is predicted by expression (17) with the relation $\epsilon =-m/2$.
In other words, these results agree with the hierarchical assumption
$\zeta\propto \xi^2$.

\section{Summary}
\label{sec:summary}
In this paper we have studied the connection between the 2PCF and the 3PCF in the non-linear gravitational clustering regime.
Under the scaling hypothesis to the bispectrum as well as the power spectrum,
we have derived a new formula regarding the 3PCF.
It is striking that the origin of the power index of the 3PCF has been
revealed.
Moreover we have checked the validity of the analytical expression
using numerical simulations in the one-dimensional Einstein-de Sitter model.

Under the scaling hypothesis, 
we have obtained the connection (\ref{eq:results}), in which
there remain two characteristic possibilities for evolution of the phases of the Fourier modes:
they remain correlated or tend to become
uncorrelated in the strongly non-linear regime.
Expression (\ref{eq:results}) states that 
the power index of the 3PCF generally consists of information from
the third-order statistics, $\epsilon$, as well as that from the 2PCF.

It is generally believed that 
the power index of the power spectrum in the non-linear
regime, $m$, is decided by the initial power spectrum. 
Combining this fact and  our formula, Eq. (\ref{eq:results}),
we can claim that the power index of the 3PCF originates from the
initial power spectrum. 
Therefore, it will be important future work to clarify the relationship 
between the power spectrum in the non-linear regime and the initial one.
Furthermore, the confirmation of the scaling of the 3PCF in three-dimensional N-body simulations is very important future work, although it is much more difficult to calculate the 3PCF in three-dimensional space than in one-dimensional.

The radical progress of recent observations
needs increasingly advanced and practical formalism for analyses 
containing the higher-order statistics.
Hence our new formula is expected to be 
a good indicator to understand the 
non-linear gravitational clustering.
For this purpose it will be a subsequent task to check if the new formula
is consistent with the BBGKY hierarchy.
In addition, it will also be important to complete the formula 
by investigating the coefficients as well as the power indices
in order to inspect the adjustment with
the ``hierarchical form'' and 
to understand the physical origin of $Q$.
We hope the new formula becomes a very useful instrument for analysis
of  large-scale structure.

\acknowledgements
We would like to thank Kei-ichi Maeda, Masahiro Morikawa, and Arihiro Mizutani for useful comments.
And we are grateful to Takahiko Matsubara and Tsutomu T. Takeuchi for many valuable comments and discussions. 
H.K. is supported by JSPS Fellowship for Young Scientists.

\appendix
\section{Bispectrum under the scaling hypothesis}
\label{sec:ap}

Firstly we decompose the absolute value of the 
Fourier spectrum of density fluctuation and the
Fourier phase into their averages and the discrepancies:
\begin{equation}
X_a= \langle X_a\rangle +\delta X_a =|\delta_{tk}|,
\end{equation}
\begin{equation}
X_b= \langle X_b\rangle +\delta X_b=|\delta_{tk'}|,
\end{equation}
\begin{equation}
X_c=\langle X_c\rangle +\delta X_c= |\delta_{-tk-tk'}|,
\end{equation}
\begin{equation}
Y= \langle Y\rangle +\delta Y=e^{i(\phi_{tk}+\phi_{tk'}+\phi_{-tk-tk'})}.
\end{equation}
Then the bispectrum is rewritten by
\begin{eqnarray}
B(tk,tk')=
\left\langle X_a X_b X_c Y\right\rangle 
&=&\left\langle 
(\langle X_a\rangle +\delta X_a)(\langle X_b\rangle +\delta X_b)
(\langle X_c\rangle +\delta X_c)(\langle Y\rangle +\delta Y)
\right\rangle 
\nonumber\\
&=&\langle X_a\rangle \langle X_b\rangle \langle X_c\rangle 
\langle Y\rangle 
+\langle X_a\rangle \langle X_b\rangle \langle X_c\rangle
\left\langle \delta Y\right\rangle +\cdots
\nonumber\\
&&+\langle X_a\rangle\langle X_b\rangle 
\left\langle \delta X_c\delta Y\right\rangle +\cdots
\nonumber\\
&&+\langle X_a\rangle 
\left\langle \delta X_b\delta X_c\delta Y\right\rangle +\cdots
\nonumber\\
&&+\left\langle \delta X_a\delta X_b\delta X_c\delta Y\right\rangle.
\end{eqnarray}
Next we rewrite these averages using 
the variances, $\sigma_a$, $\sigma_b$, 
$\sigma_c$ and $\sigma_y$ and 
the correlation coefficients,
$r_{ab}$, $r_{ay}$, $r_{abc}$, $r_{aby}$ and $r_{abcy}$,
\begin{eqnarray}
 \langle X_a\rangle &=&\sigma _a,\nonumber\\
\left\langle \delta X_a^2\right\rangle&=&s^2\sigma _a^2, \nonumber\\
\left\langle \delta Y^2\right\rangle&=&\sigma _y^2,\nonumber\\
\left\langle \delta X_a\delta X_b \right\rangle 
&=& s^2\sigma _a\sigma _b r_{ab},
\nonumber\\
\left\langle \delta X_a\delta Y \right\rangle 
&=& s\sigma _a\sigma _y r_{ay},
\nonumber\\
\left\langle \delta X_a\delta X_b\delta X_c \right\rangle 
&=& s^3\sigma _a\sigma _b\sigma _c r_{abc},
\nonumber\\
\left\langle \delta X_a\delta X_b\delta Y \right\rangle 
&=& s^2\sigma _a\sigma _b\sigma _y r_{aby},
\nonumber\\
\left\langle \delta X_a\delta X_b\delta X_c\delta Y \right\rangle 
&=& s^3\sigma _a\sigma _b\sigma _c\sigma _y r_{abcy},
\end{eqnarray}
where $s$ is a constant of order $1$. Then 
\begin{eqnarray}
\left\langle X_a X_b X_c Y\right\rangle 
&=&\sigma _a\sigma _b\sigma _c\langle Y\rangle
+s\sigma _a\sigma _b\sigma _c\sigma _y(r_{ay}+r_{by}+r_{cy})
+s^2\langle Y\rangle\sigma _a\sigma _b\sigma _c(r_{ab}+r_{bc}+r_{ca})
\nonumber\\&&
+s^2\sigma _a\sigma _b\sigma _c\sigma _y(r_{aby}+r_{bcy}+r_{cay})
+s^3\langle Y\rangle\sigma _a\sigma _b\sigma _cr_{abc}
+s^3\sigma _a\sigma _b\sigma _c\sigma _yr_{abcy}.
\label{A7}
\end{eqnarray}
 Now we assume that these constants obey power laws as follows:
\begin{eqnarray}
\sigma_a\propto t^{m/2}, \quad
\langle Y\rangle\propto t^{-\alpha},
\label{A8}
\end{eqnarray}
\begin{eqnarray}
\sigma_y\propto t^{-\beta},\quad
r_{ab} \propto t^{-\gamma _2},\quad r_{ay} \propto t^{-\gamma _{2y}}, \quad
r_{abc} \propto t^{-\gamma _{3}},\quad r_{aby} \propto t^{-\gamma _{3y}},\quad
r_{abcy} \propto t^{-\gamma _{4y}}.
\label{A9}
\end{eqnarray}
Substituting (\ref{A8}) and (\ref{A9}) into (\ref{A7}), we find 
\begin{eqnarray}
\left\langle X_a X_b X_c Y\right\rangle 
&\propto&t^{3m/2-\alpha}+t^{3m/2-\beta-\gamma_{2y}}+t^{3m/2-\alpha-\gamma_{2}}
+t^{3m/2-\beta-\gamma_{3y}}+t^{3m/2-\alpha-\gamma_{3}}
+t^{3m/2-\beta-\gamma_{4y}}.\nonumber\\
\end{eqnarray}
In order for the 3PCF not to diverge, these scaling factors, 
$\alpha$, $\beta$, and $\gamma$
should be non-negative values. 
Then the bispectrum can be approximated as
\begin{equation}
\langle X_{a}X_{b}X_c Y\rangle \propto t^{3m/2-\epsilon},
\end{equation}
where $\epsilon$ is defined as
\begin{equation}
\epsilon\equiv min(\alpha,\beta+\gamma_{2y},\beta+\gamma_{3y},
\beta+\gamma_{4y}).
\end{equation}

\end{document}